# When Regression Verification Meets CEGAR

Fei He
School of Software, Tsinghua University
Haidian, Beijing, China
hefei@tsinghua.edu.cn

Qianshan Yu
School of Software, Tsinghua University
Haidian, Beijing, China
yuqianshan@foxmail.com

Liming Cai
School of Software, Tsinghua University
Haidian, Beijing, China
limingcai0101@yeah.net

## ABSTRACT

Software systems evolve throughout their life cycles. Many revisions are produced over time. Model checking each revision of the software is impractical. Regression verification suggests reusing intermediate results from the previous verification runs.

This paper proposes a fully automatic regression verification technique in the context of CEGAR. Procedure summaries, which describe the input/output behaviors of a procedure, are proposed as the intermediate results to be reused. Procedure summaries are reasonably small to store, technically easy to process, and do not require much extra computation effort to be reused. Reusing procedure summaries saves much analysis effort on the corresponding procedures. By combining regression verification and CEGAR, we propose a technique that is able to reuse procedure summaries across different abstract precisions and different program revisions.

We performed extensive experiments on a large number of industrial programs (534 revisions of 89 Linux kernel device drivers). The results show that our approach can significantly improve the performance of regression verification.

## 1 INTRODUCTION

Along with the widespread use of software in our daily life, there is a growing concern for software reliability. At the same time, market pressure demands quick product introductions. The software companies are required to introduce new features to their software products in shorter release cycles. Since errors may be introduced with new features, each revision of these software products needs be formally verified.

Model checking [16] is one of the most successful techniques in formal verification. However, it is still very time-consuming. Verifying every revision of the software is impractical. Inspired by the regression testing [20, 31], researchers in formal verification community propose the technique of regression verification [2, 10, 11, 19]. Taking into consideration that many intermediate results are produced during the verification, and computing these results is costly, regression verification aims to make use of these intermediate results in the verification of new program revisions.

Different intermediate results have been proposed for reuse, including abstract precisions, state-space graph, constraint solver solutions, and interpolation-based procedure summaries. Beyer et al. [10] proposed to record the final abstract precision in the previous verification run, and reuse it in the current verification. Henzinger et al. [23] proposed to reuse state-space graph for incremental checking of temporal safety properties. Visser et al. [35] noticed the important role of constraint solving in software verification, and proposed to reuse the constraints solving results. Sery et al. [32] proposed to reuse the interpolation-based procedure summaries in incremental upgrade checking.

This paper proposes to use state-based procedure summaries as the intermediate verification results. A state-based procedure summary is a pair of input and output states, describing the observable behaviors of the procedure. Remark that the state-based procedure summaries is a different type of intermediate results to the interpolation-based summaries [1]. In the following, unless otherwise specified, by procedure summary we mean the state-based summary.

The state-based procedure summaries can be automatically generated by most of the interprocedural analysis techniques [30, 33]. Generating these procedure summaries does not need much additional computation effort. Recall that a state-based procedure summary is a pair of input and output states. They are efficient to store, easy to process, and do not require much extra computation effort before they can be reused. More importantly, reusing state-based procedure summaries avoids re-analysis of the same procedures, and thus significantly reduces the verification effort for the next revision. State-based procedure summaries, therefore, are a good choice for reusing in regression verification.

We consider regression verification in the context of counterexample guided abstraction refinement (CEGAR) [15]. Techniques for representing, processing and reusing state-based procedure summaries need be adapted to the CEGAR framework (see Section 4 and 5). A lazy counterexample analysis technqiue is further proposed to address the effectiveness issuer of summary reuse (see Section 6). To the best of our knowledge, we are the first to consider the reuse of state-based procedure summaries in the context of CEGAR. Considering that CEGAR is a widely-adopted technique in software verification [4, 6, 9, 13, 22], our approach can be applied to most state-of-the-art software verifiers.

---
[1]Their differences will be discussed in Section 8.

We implemented our approach on top of CPAchecker[9], and performed extensive experiments on a large set of industrial programs. Up to 534 revisions of 89 Linux kernel device drivers are used in experiments. Experimental results show very promising performance of our approach. In comparison to the standalone verification, our summary reuse technique saves 31.4% of verification time. We also test the impact of summary reduction and lazy counterexample analysis to regression verification. Results show that these two techniques can significantly improve the performance of regression verification.

The main technical contributions of this paper are summarized as follows:

- We identify the value of state-based summaries as intermediate results to be reused in regression verification.
- We propose a fully automatic regression verification technique in the context of counterexample-guided abstraction refinement, with which procedure summaries can be efficiently reused across different abstract precisions and different program revisions.
- The counterexample analysis is nontrivial in the context of regression verification and CEGAR. We propose a lazy analysis technique to improve the effectiveness of summary reuse.
- We implemented our approach in the software verification tool CPAchecker. Experimental results show promising performance of our approach.

The remainder of this paper is organized as follows. Section 2 introduces necessary backgrounds. Section 3 illustrates our approach by a toy example. Section 4 presents techniques for representing and processing summaries in the context of CEGAR. Section 5 presents our regression verification framework. Section 6 exhibits the lazy counterexample analysis technique. Section 7 reports evaluation results on our approach. Section 8 discusses related work and Section 9 concludes this paper.

## 2 BACKGROUND
### 2.1 Abstraction and Refinement

Abstraction plays a central role in software verification. Abstraction omits details of the system behaviors, resulting in a simpler model. We call the model before and after abstraction the *concrete* and *abstract* model, respectively. An abstraction is *conservative* [16] if it does not omit any behavior of the concrete model. Conservative abstraction guarantees that the properties (more precisely, the $ACTL^*$ properties [16]) established on the abstract system also hold on the concrete system. The reverse, however, is not guaranteed: if the abstract model falsifies the property, the concrete model does not necessarily falsify this property, too.

The *abstract precision* [10] (for short, *precision*) defines the level of abstraction of an abstract model. The precision must be at a proper level. A too-coarse precision may fail to verify the property, a too-fine precision, however, may lead to state space explosion. Finding a proper precision appears to require ingenuity.

*Counterexample-guided abstraction refinement* [15] provides a framework for automatically finding proper precisions. Starting from an initial abstract precision, it iteratively checks if the corresponding abstract model satisfies the desired property. If the property is satisfied, it must also hold on the concrete model, the algorithm terminates and reports "safe". Otherwise, the checker returns a path on the abstract model that falsifies the desired property. The algorithm then checks if the returned path is valid on the concrete model or not. If it is, the algorithm finds a real bug, it thus terminates and reports "unsafe". Otherwise, the precision is too coarse, and needs be refined with the counterexample. Then the above process repeats, until either "safe" or "unsafe" is reported.

The abstract precision is not necessary to be the same throughout the program [24]. To simplify the discussion, we assume in this paper that the abstract precisions are defined at the procedure level, i.e., each procedure is associated with an unique abstract precision. Note that this is not a limitation of our approach, which can be applied to other settings of abstract precisions.

### 2.2 Software Verification

Model checking and program analysis are two major approaches for software verification. Comparing these two techniques, model checking is more precise with fewer false positives produced, while program analysis is comparatively more efficient and can be applied to more programs. An increasing tendency to software verification is to integrate these two techniques together [8], to get a good balance between accuracy and efficiency.

*Control flow automata* (CFA) are adopted in many software verification techniques (for example, BLAST and CPAchecker) for representing programs. Let $X$, $\mathcal{L}$ and $Op$ be the sets of variables, locations and operations in a program $P$, respectively. The CFA of $P$ consists of the set $\mathcal{L}$ of locations and the set $G \subseteq \mathcal{L} \times Op \times \mathcal{L}$ of control flow edges. A *state* of $P$ is a pair $(l, u)$, where $l \in \mathcal{L}$ is a program location and $u$ is a valuation to $X$.

*Predicate abstraction* is a popular abstraction technique for software verification. The abstract precision $\lambda$ is a set of predicates. All predicates are defined over the program variables. This analysis tracks values of predicates in $\lambda$ on each abstract state. Formally, an abstract state for predicate analysis is a pair $(l, s)$ where $l$ is a program location and $s$ an abstract valuation that assigns Boolean values to all predicates in $\lambda$. In the remainder of the paper, we also call $s$ an *abstract data state*. There exists other abstract domains, like interval abstraction, octagon abstraction, etc. Our technique can easily be extended to these abstract domains.

*Interprocedural analysis* deals with programs with multiple procedures. A simple way for interprocedural analysis is to inline a copy of the callee procedure at each of its callsites. The inlining technique is, however, very expensive and may lead to context explosion for recursive procedures. Another interprocedural analysis technqiue is by use of procedure summaries [30, 33]. A procedure summary (or shortly, a summary) describes the input/output behaviors of a procedure. This technique plugs summaries in at each callsite of the procedure. With this technique, we avoid reanalyzing the procedure body at each of its call sites, the efficiency is thus improved.

```
main(){
    …
1:  if (flag) {
2:      y = inc(x, flag);
3:      if (y <= x) ERROR;
    } else {
4:      z = inc(x, flag);
5:      if (z >= x) ERROR;
    }
    …
}

inc(int a, int sign){
    int rv;
6:  if (sign)
7:      rv = a + 1;
    else
8:      rv = a - 1;
9:  return rv;
}
```

Figure 1: An example program

## 3  A MOTIVATING EXAMPLE

Figure 1 shows a simple program we designed to illustrate the basic process of summary reuse in the context of CEGAR. This program consists of two procedures: *main* and *inc*. The *inc* procedure takes two input parameters: *a* and *sign*, and outputs either $a+1$ (if $sign \neq 0$), or $a-1$ (if $sign = 0$). The *inc* procedure is invoked twice in the *main* procedure. We want to verify that the "ERROR" statements (at line 3 and 5) are not reachable in any execution of this program.

Assume the current abstract precision for the *inc* procedure contains two predicates, i.e., $sign = 0$ and $rv > a$. Only values of these two predicates need be tracked in the predicate analysis of the *inc* procedure. At the first call (at line 2) to the *inc* procedure, $flag \neq 0$, and thus the predicate $sign = 0$ is false. The invoked *inc* procedure therefore takes the *true* branch (line 7), and at the return point the predicate $rv > a$ holds. One summary is concluded after this procedure call, written

$$\langle \neg(sign = 0), \neg(sign = 0) \wedge rv > a \rangle, \quad (1)$$

meaning that if the input state of the *inc* procedure satisfies the left formula (called the input condition), the output state of the *inc* procedure must satisfy the right formula (called the output condition). Similarly, at the second call (line 4) to the *inc* procedure, the invoked *inc* procedure takes another branch and concludes

$$\langle sign = 0, sign = 0 \wedge \neg(rv > a) \rangle. \quad (2)$$

Above two formulas summarize the input/output behaviors of the *inc* procedure at the given abstract level.

For more information on the summary generation of interprocedural analysis, please refer to [21, 37].

### 3.1  Reuse Across Invocations

If there is another call to the *inc* procedure in the remainder of the program, without entering the procedure's body, we can immediately determine its output state by testing $\neg(sign = 0)$ on its input state.

```
main() {
    …
1:  if (flag)
2:      y = inc(x, flag);
    else
3:      x = inc(y, flag);
4:  if (x > y) ERROR;
    …
}
```

Figure 2: The updated *main* procedure

Summary reuse across different invocations has already been supported by existing interprocedural analysis techniques [30, 33]. A *summary cache* is maintained to keep all procedure summaries generated so far. In case of a procedure invocation, the verifier attempts to seek in the cache for a summary that meets the current input state. Only when no applicable summary exists in the cache, do we need to analyze the procedure.

### 3.2  Reuse Across Programs

Consider now the original program evolves to a new version, where the *main* procedure changes, while the *inc* procedure does not. The updated *main* procedure is shown in Fig. 2. Apparently, this new program must be re-verified to guarantee its correctness. Traditional interprocedural techniques [30, 33] verify this new program from scratch, which apparently is not an efficient way.

Note that there are many commonalities between two consecutive revisions. The summaries generated for the older revision have many possibilities to be reused in the regression verification. These summaries contain important information about the verification. Reusing them can significantly reduce the consequent verification effort.

Considering the example, since the *inc* procedure remains the same in the new revision, the previously generated summaries (1) and (2) can be applied to the verification of the new revision. At line 2 of the new *main* procedure, without entering the *inc* procedure, we can directly get that $y > x$ by applying (1) [2]. Similarly, at line 3, we get that $\neg(x > y)$ by applying (2). In either case, the condition $y < x$ does not hold, we thus conclude the new program is safe. Remark that in the above regression verification, we need not to look into the body of the *inc* procedure.

There are some issues that must be addressed for reusing summaries across different program versions:

- Abstraction issue: Note that the regression verification involves a new round of CEGAR. Abstract precision changes among different iterations of CEGAR. Summaries generated in the previous verification run may not be applicable to the current verification. Reconsidering the above example, if the current abstract precision does not contain the predicates $rv > a$ and $rv < a$, the summaries (1) and (2) can not be applied.
- Counterexample issue: In the framework of CEGAR, each time the verifier returns a counterexample, one needs to check if this counterexample corresponds to a real path or not. To validate a counterexample, the involved procedure calls need be expanded to the inner paths in the corresponding procedures. This task is nontrivial and may lead the summary reuse much less useful for regression verification (see Section 6). A smarter counterexample analysis technique is in demand.

---

[2] Assume the predicates $y > x$ and $x > y$ are included in the abstract precision of the new *main* procedure.

## 4 ABSTRACT SUMMARIES

To adapt to CEGAR, the summary definition needs to be extended with the abstract precision. In this section, we first introduce the definition of abstract procedure summaries, then discuss their representations, and finally propose an optimization technique for summary reduction.

### 4.1 Definitions

Let $p$ be a predicate, a *predicate literal* of $p$ is either $p$ or $\neg p$. Recall that an abstract precision $\lambda$ is a set of predicates. An *abstract data state* is a valuation to all predicates in $\lambda$. An abstract data state can be represented as a complete conjunction of predicate literals, such that each predicate in $\lambda$ occurs in the conjunction in either positive or negtive form. In contrast, a partial conjunction of predicate literals represents a set of abstract data states.

Considering the motivating example, assume the current abstract precision consists of two predicates, i.e., $sign = 0$ and $rv > a$. Positive and negative forms of these two predicates are all predicate literals. The formula "$sign = 0 \land \neg(rv > a)$" is a full conjunction of predicate literals in the precision, representing an abstract data state. The formula "$sign = 0$" is a partial conjunction of predicate literals, representing a set of abstract data states (the predicate $rv > a$ can be either true or false).

*Definition 4.1.* A *summary* of a procedure $f$ with respect to an abstract precision $\lambda$ is a pair $\langle \phi_{in}, \phi_{out} \rangle$, where $\phi_{in}$ and $\phi_{out}$ are conjunctions of predicate literals in $\lambda$, representing the input and output conditions of $f$, respectively.

A summary $\langle \phi_{in}, \phi_{out} \rangle$ indicates that if the input state of $f$ satisfies $\phi_{in}$, its output state must satisfy $\phi_{out}$. In our implementation, we require $\phi_{in}$ to be a complete conjunction of predicate literals and $\phi_{out}$ a "partial" one. This is accordant with the process of intra-procedural analysis [9], since from a single abstract state, there usually can be more than one reachable abstract states at the exit point of the procedure.

All procedure summaries are kept in a *summary cache*. During the program analysis, if a procedure call statement is encountered, the verifier seeks in the cache for a proper summary. Let the called procedure be $f$, the current abstract precision be $\lambda$, and the current abstract data state be $s$. A summary of $f$ is *applicable* to this procedure call, if

(1) it is with the same abstract precision $\lambda$, and
(2) its input condition $\phi_{in}$ is satisfied by $s$.

If there exists an applicable summary, the cache returns $\phi_{out}$ of the summary to the verifier. Otherwise, it returns "NULL". In the latter case, the verifier needs to conduct a heavy intra-procedural analysis to compute the output.

Note that testing if the input condition $\phi_{in}$ of a summary is satisfied by the current abstract state requires a satisfiability checking.

In the worst case, the number of satisfiability checkings are up to the total number of summaries in the cache.

### 4.2 Summary Representation

Note that a large number of satisfiability checking is required to find an applicable summary. To improve the efficiency of summary retrieval, we use ordered binary decision diagrams (OBDDs) to represent procedure summaries.

OBDD is a compact and canonical representation of Boolean functions. If two Boolean functions are equivalent, their OBDDs must be isomorphic. The canonicality of OBDD leads to a very efficient algorithm for checking equivalence of two Boolean functions [12]. Moreover, many modern OBDD packages (for example, the CUDD package [34]) realizes the shared OBDD technique, with which the equivalence checking of two Boolean functions can be finished in constant time [34].

For each procedure summary, two OBDDs that represents $\phi_{in}$ and $\phi_{out}$, respectively, are stored. Construction of these two OBDDs may take some time. However, it is worthwhile, since these OBDDs may later be tested many times during the verification. With the OBDD representation, each test requires only a constant time.

### 4.3 Summary Reduction

A procedure summary may contain redundant information. This redundant information is useless, and may decrease the hit rate of the summary cache. In the following, we discuss how to remove redundant information from summaries.

A number of variables are accessed in the body of a procedure, including formal parameters, local variables and global variables. Let $V_f$ be the set of all variables accessed by the procedure $f$. A predicate is *relevant* to $f$ if it contains at least one variable in $V_f$. A predicate is *irrelevant* to $f$, if all variables occurred in the predicate are irrelevant to the procedure. For example, $a$ and $sign$ are formal parameters, and $rv$ is a local variable of the *inc* procedure, thus the predicates $sign = 0$, $rv > a$ are both relevant, while the predicates $flag = 0$ and $x > 0$ are not.

The irrelevant predicates have nothing to do with the procedure, and can be safely removed from this procedure's summaries. Recall that each summary is associated with an abstract precision. To reduce a summary, the corresponding abstract precision needs also be reduced. Let $\langle \phi_{in}, \phi_{out} \rangle$ be a summary with respect to an abstract precision $\lambda$. Recall that $\lambda$ is a set of predicates; $\phi_{in}$ and $\phi_{out}$ can be considered as two sets of predicate literals. The summary reduction proceeds as follows: identifies and removes all irrelevant predicates in and from $\lambda$; then removes all occurrences of these irrelevant predicates in both $\phi_{in}$ and $\phi_{out}$.

Note that the above summary reduction procedure is not *complete*. It cannot guarantee removing all redundant information. For a more powerful method, one may refer to the method of quantifiers elimination. However, our technique is much more efficient than that of quantifiers elimination. Summary reduction is just an optimization, we are more interested in efficiency here.

Note that at each callsite of a procedure, the current abstract precision and the current input state need also be reduced. Then with these reduced abstract precision and reduced input state, the verifier look applicable summaries up in the cache.

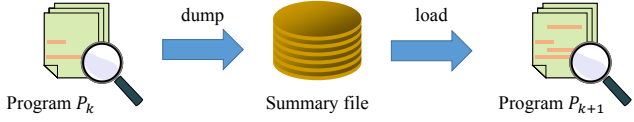

Figure 3: Summary reuse across different program versions

Consider the program example in Fig. 1, assume the abstract precision of the *inc* procedure is $\{y > 0, sign = 0, rv > a\}$, a possible summary of the procedure is

$$\langle y > 0 \land \neg(sign = 0), y > 0 \land \neg(sign = 0) \land rv > a \rangle. \quad (3)$$

Assume the *inc* procedure is called again at a subsequent position, and assume the input state at that callsite is $\neg(y > 0) \land \neg(sign = 0)$. Apparently, the input condition of (3) is not satisfied by this input state. However, if we look into this summary, we find that the predicate $y > 0$ is indeed irrelative to the *inc* procedure. By removing this predicate, we get a reduced summary

$$\langle \neg(sign = 0), \neg(sign = 0) \land rv > a \rangle.$$

This reduced summary is now applicable to the second invocation. Removing irrelevant predicates can thus improve the hit rate of summaries .

## 5 SUMMARY REUSE FOR REGRESSION VERIFICATION

Summaries convey important information about the verification. To reuse summaries in regression verification, we need to dump all summaries to an external file, called the summary file, after each verification run; and load the previously-generated summaries from the summary file to initialize the current summary cache, before a regression verification starts (Fig. 3). In this way, the important information stored in the procedure summaries is shared across different program versions.

### 5.1 Summary Dumping

Summary dumping is quite simple. When a verification finishes, the verifier dumps all entries of the summary cache to an external file. Remark that this operation is performed in the end of the verification, and does not intervene to the verification process.

We define a simple text-based format for the summary file. A fragment of the summary file for the program example is shown in Fig. 4, where the first line declares these summaries belong to the *inc* procedure. Note that a procedure identifier consists of its name and its type signature. Summaries of a procedure are grouped by abstract precisions. Lines 2 to 3 of Fig. 4 declare an abstract precision, while lines 4 to 5, and 6 to 7 declare two summaries with respect to this precision. The format for declaring predicates and formulas follows the SMT-LIB standard [3].

### 5.2 Summary Loading

Before the regression verification starts, we load the previously-generated summaries from the summary file and then reuse them in the current verification.

```
1:  inc(int,int)
    {
      …
2:    (assert (= sign 0))
3:    (assert (> rv a))
      {
4:      (assert(not (= sign 0)))
5:      (assert(and (not (= sign 0)) (> rv a )))
      }
      {
6:      (assert(= sign 0))
7:      (assert(and (= sign 0) (not (> rv a))))
      }
      …
    }
```

Figure 4: A summary example in text format

Due to program changes, not all summaries of the old program can be reused in the current verification. Let $P$ and $P'$ be two versions of a program. Let $f$ be a procedure in $P$, and $f'$ its updated version in $P'$. Procedure summaries of $f$ can be reused in the regression verification, if and only if $f$ and $f'$ are semantically equivalent.

However, checking semantic equivalence of each procedure is very time-consuming. We choose to perform syntactic checking instead. More specially, we syntactically compare $P$ and $P'$ to find a set $\tau_P$ of *syntactically unchanged* procedures. Note that a procedure may call other procedures in its body. If the callee procedure changes, even the caller procedure remains syntactically the same, its semantics changes. We thus find a subset $\tau_P^*$ of $\tau_P$, such that for any procedure $f \in \tau_P^*$, all its callee procedures are also in $\tau_P^*$. Note that this is a recursive definition. Then we have the following lemma.

LEMMA 5.1. *For any procedure in $\tau_P^*$, its two versions in $P$ and $P'$ are semantically equivalent.*

Compared to semantic equivalence checking, our syntactic checking is less precise. There may exist procedures that are semantically equivalent but syntactically different. Summaries of such procedures will be abandoned by our technique. Note that our syntactic checking is much more efficient. We indeed trade the maximal reuseability for efficiency.

We use the set $\tau_P^*$ to filter summaries in the process of summary loading. Only summaries that belong to some procedure of $\tau_P^*$ are kept, and all others are directly abandoned. According to Lemma 5.1, all kept summaries belong to semantically equivalent procedures, and can be safely reused in the regression verification. Remark that the computation of $\tau_P^*$ and loading summaries can be realized as a preprocessing step of the regression verification, and does not intervene to the real verification process.

## 6 LAZY COUNTEREXAMPLE ANALYSIS FOR REGRESSION VERIFICATION

In the context of regression verification and CEGAR, the counterexample analysis is nontrivial. We explain in the following the problem and then present our solution.

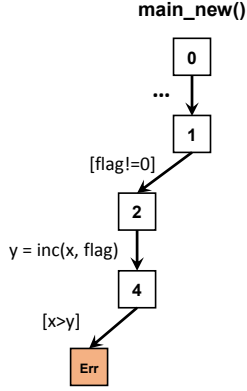

Figure 5: A counterexample

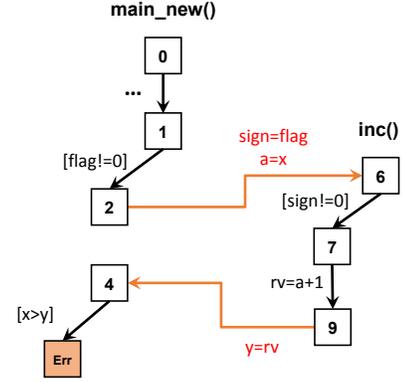

Figure 6: The expanded counterexample

## 6.1 The Problem

If the property cannot be verified with the current abstract precision, the verifier returns a counterexample. We need to check if this counterexample corresponds to a real bug or not.

Fig. 5 shows a counterexample path, obtained by regression verification of the updated *main* procedure in Fig. 2. Each node on the path represents an abstract state. Recall that an abstract state consists of a program location and an abstract data state. In the figure, only the program locations are labeled, whereas the abstract data states, for simplicity reason, are skipped.

Note that there is a procedure call between $l_2$ and $l_4$ in the counterexample. To validate this counterexample, the intra-procedural path of the *inc* procedure needs be restored. Let $s_2$ and $s_4$ be the corresponding abstract data states at $l_2$ and $l_4$, respectively. Assume in the verification step, the abstract data state $s_4$ is computed by intra-procedural analysis on the *inc* procedure. After the intra-procedural analysis, the reached states from $s_2$ of the *inc* procedure are computed and cached in memory (for example, in the CPAchecker [9]). Now in the counterexample analysis step, with this reached set, the intra-procedural path of *inc* from $s_2$ to $s_4$ can be easily restored.

Assume in the verification step, the output state of the *inc* procedure at $s_2$ is obtained from a cached summary that was generated in the previous verification. Then in the counterexample analysis, there is no information for the reached set of the *inc* procedure, We have to rely on the costly intra-procedural analysis to build the inner path in *inc* from $s_2$ to $s_4$. That is to say, the saved intra-procedural analysis gets back in the counterexample analysis. The benefits of summary reuse are thus annulled to some extent by the counterexample analysis.

*Definition 6.1.* A procedure call operation is called a *hole* if the corresponding reached set of this procedure is not available.

## 6.2 Our Solution

Let $P$ be a program. An *abstract path* $\pi$ is an alternating sequence of abstract states and operations, i.e.,

$$\pi = (l_0, s_0) \xrightarrow{op_0} (l_1, s_1) \xrightarrow{op_1} \ldots \xrightarrow{op_{n-1}} (l_n, s_n).$$

The path $\pi$ is called a *CFA path* if $l_0$ is the entry location of the program, and for each $i$ with $0 \leq i < n$ there exists a CFA edge $g = (l_i, op_i, l_{i+1})$. In other words, a CFA path represents a syntactical walk through the CFA. A *counterexample* is a CFA path that ends at the error location $l_{err}$.

A path may contain procedure call operations. Replacing the procedure call operation with the corresponding intra-procedural path is called an *expansion*. In the following, we assume all procedure call operations, except of holes, have already been expanded. For example, the expansion of the counterexample in Fig. 5 is shown in Fig. 6. We call a path *holeless* if it contains no hole.

Define the *strongest post-condition operator SP* as follows: for a formula $\varphi$ and an operation $op$ ($op$ is not a hole), $SP_{op}(\varphi)$ represents the set of data states that are reachable from any of the states that satisfy $\varphi$ after the execution of $op$. The *concrete semantics* of a holeless path is the successive applications of the $SP$ operator to operations of $\pi$, i.e., $SP_\pi(\varphi) = SP_{op_n}(\ldots(SP_{op_0}(\varphi))\ldots)$.

*Definition 6.2.* A holeless path $\pi$ that starts from the abstract state $(l, s)$ is *feasible* if $SP_\pi(s)$ is satisfiable.

Let $H$ be the set of holes on the path $\pi$. These holes split $\pi$ into $|H| + 1$ path segments. Each of these segments is a holeless path. The concrete semantics of $\pi$ is defined as the conjunction of the concrete semantics of these segments.

Theorem 6.3. *Let $\pi$ be a path with holes $H$, and $\Pi$ the set of segments of $\pi$ split by $H$,*

(1) *if any segment in $\Pi$ is infeasible, $\pi$ is infeasible; and*
(2) *if all segments in $\Pi$ are feasible, $\pi$ is, however, not necessary to be feasible.*

For the latter case, if all segments of $\pi$ are feasible, we call the path $\pi$ is *separately feasible*. The above theorem states that the separately feasibility cannot prove the feasibility of the whole path. This is obvious since the concrete semantics of holes are not taken into consideration in the separately feasibility.

Consider the counterexample in Fig. 5, assume the procedure call operation at $l_2$ is a hole. It splits the path into two segments, i.e.

$$\sigma_1 = (l_0, s_0) \xrightarrow{\cdots} (l_1, s_1) \xrightarrow{[flag!=0]} (s_2, s_2),$$

$$\sigma_2 = (l_4, s_4) \xrightarrow{[x>y]} (l_{err}, s_{err}).$$

This counterexample is infeasible if either $SP_{\sigma_1}(true)$ or $SP_{\sigma_2}(s_4)$ is unsatisfiable.

Our *lazy analysis* algorithm is shown in Alg. 1. The basic idea is to expand holes only when needed. In this way, we avoid unnecessary hole expansions. The main body of the algorithm is a *while* loop. In the beginning of each iteration of the loop, the algorithm checks whether the current path is holeless (*isHoleless*($\pi$)) or not, and whether the current path is infeasible (*isInfeasible*($\pi$)) or not. If both checks returns *false*, the loop continues by expanding one hole in $\pi$. Otherwise, the current path $\pi$ must be either infeasible or holeless. For the former case, the algorithm returns $\pi$; and for the latter case, it returns "*unsafe*".

**Input:** A finite abstract path $\pi$
**Output:** The expanded path of $\pi$ if it is infeasible; or "*unsafe*" if it corresponds to a real path.
**while** ¬(*isHoleless*($\pi$) ∨ *isInfeasible*($\pi$)) **do**
　　let $h$ be a hole in $\pi$;
　　$\pi \leftarrow expandHole(\pi, h)$;
**end**
**if** *isInfeasible*($\pi$) **then return** $\pi$ ;
**else return** *unsafe* ;

**Algorithm 1:** *lazyAnalysis*($\pi$)

A *brute-force* approach for counterexample analysis is to directly expand all holes of the counterexample, and then check its feasibility. Compared to our lazy approach, the brute-force approach needs only one feasibility checking. However, our approach is still more beneficial. Firstly, with our lazy approach, the computational efforts for unnecessary hole expansions (which in many cases are time-consuming) are saved. Secondly and more importantly, the returned path by our approach is often much shorter than the fully expanded one. Note that the refinement is a heavy step in CEGAR. With a shorter counterexample, the computation efforts for the refinement (for example, the interpolation-based refinement [28]) can often be significantly reduced.

Our lazy analysis algorithm can be easily adapted to the existing CEGAR framework (for example, CPAchecker [9]). When the verifier in the existing framework returns a counterexample, our algorithm is applied to check if this counterexample is spurious or not. In case of spurious counterexample, our algorithm returns a (partially) expanded path, and gives it to an existing refiner in the framework. The returned path by our algorithm may contain holes. Treating these holes as value assignments, these paths can be directly processed by most of the existing refinement techniques, for example, the interpolation-based refinement [28].

## 7 EXPERIMENTAL EVALUATION

We implemented our regression verification technique on top of CPAchecker [9]. CPAchecker provides a configurable framework for software verification. Predicate abstraction and counterexample guided abstraction refinement are supported in CPAchecker. We need to add supports for summary dumping, summary loading and lazy counterexample analysis.

To evaluate the effectiveness and efficiency of our summary reuse technique to the regression verification, we perform extensive experiments on a large set of industrial programs. We take device drivers from Linux kernel as the benchmark. All drivers are extracted from the "SystemsDeviceDriversLinux64ReachSafety" category of the 6nd International Competition on Software Verification (SV-COMP'17) [5]. We choose programs according to the following two criteria: (1) the lines of code is no less than 1000, and (2) at least one refinement is required for verifying this program using CPAchecker. In this way, we omit the trivial programs and those that CEGAR is not needed. Moreover, we limit our selection to drivers from Linux 3.4 kernel and with the mutex lock/unlock specifications [10].

We use a higher-order mutation tool MiLu [26] to generate new revisions of these drivers. Up to 5 new revisions are randomly generated for each device driver. Counting up the initial revision, there are 6 revisions for each driver. In total, we prepared a benchmark set with 534 revisions of 89 device drivers.

For each device driver, the initial revision is verified from scratch. Each mutated revision corresponds to a regression verification task. Verification tasks are performed in an incremental way. Procedure summaries of the former revision are reused in the next revision. In total, there are 534 verification tasks, among which 445 are regression verification tasks.

All experiments are performed on a machine with a 3.6GHz 8 Core CPU and 16GB RAM. We use Ubuntu 16.04 (64-bit) with Linux 4.13 and jdk1.8.0. We use the *predicateAnalysis-bam* configuration for CPAchecker. Each verification run is limited to 90 seconds of run time, 11GB of Java heap size and 4 CPU cores.

### 7.1 Overall Results

We compare the performance of regression verification with and without summary reuse in this experiment. For simplicity, in the following, we refer to our approach as *summary reuse*, and the approach without summary reuse as *standalone*.

Experimental results are listed in Table 1. We restrict this table to the 35 best and 5 worst cases out of the total of 89 device drivers (sorted by the *Speedup* column). The first column (*Device Driver*) lists the device drivers' names. The second column (*Rst*) shows the verification results, where the first and the second number indicate the amount of "safe" and "unsafe" results, respectively. The third column (*Loc*) shows the lines of code for the initial version of each device driver. The fourth column ($T_{1st}$) lists the verification time for the first version of each device driver. Recall that this is not a regression verification task. $T_{1st}$ gives us information on the complexity of verifying each device driver.

The following two column assemblies report the experimental results of *standalone* and *summary reuse*, respectively. For each approach, we report the total verification time ($T_{Total}$), the verification time on the refinement step ($T_{Ref}$) and the number of refinements (*#Ref*). All results about time are expressed in seconds. Note that the verification time on all steps of CEGAR, and the time for processing and accessing procedure summaries are all counted in $T_{Total}$. For *summary reuse* approach, we also report the maximal size of summary files (*MaxFSize*) among all revisions of each device driver. The file size is expressed in Kilobyte. The last column shows the time speedup of *summary reuse* over *standalone*. Recall that there are 5 mutated revisions for each device driver. Each row in the table sums up results of all regression verification tasks of the corresponding driver. The last (*Total*) row sums up results of all rows in the table (except for the *MaxFSize* column that takes the maximal value).

From Table 1, we observed that *summary reuse* outperforms *standalone* in most of cases. More specially, among all 89 drivers, *summary reuse* wins in 72 drivers. Our approach is a little bit slower for 17 drivers. Looking at the 5 worst cases, their first-version verification time are all less than 0.5 seconds. It is understandable that on trivial verification tasks, our summary reuse approach may not get the best performance. In total, the *standalone* approach finishes all 445 regression verification tasks in 670.4 seconds, while our *summary reuse* approach finishes in 460.2 seconds. The overall time speedup of our approach is 31.4%.

Comparing the numbers of refinements (*# Ref*) required by *summary reuse* and *standalone* for each device driver, there is no obvious pattern. Recall that with our approach the counterexample is analyzed in a lazy fashion. The path given to a *refiner* may not be as same as that using the *standalone* approach. Hence, the whole refinement process may differ. However, if we look at the refinement time ($T_{Ref}$), we find that our approach is noticeably superior to *standalone*. Using our approach, the total refinement time is decreased from 97.4 to 42.7 seconds (about 56%). These results evidence the effectiveness of our lazy counterexample analysis technique.

Let us look at the *MaxFSize* column. The maximal size of summary files among all device drivers is 774.0KB (for most drivers, less than 500KB). The added overhead by our approach in storage is acceptable.

### 7.2 Impact of Summary Reduction

The second experiment tests the impact of summary reduction to our regression verification technique. We compare the performance of our approach with and without the summary reduction.

The results are plotted in Fig. 7, where each point represents a regression verification task, with the *X* and *Y* axes representing the regression verification time using our approach with and without summary reduction, respectively. There are totally 445 points in the figure. Timeout cases (exceeding 90 seconds) are projected to the corresponding axis. Note that both *X* and *Y* axes are logarithmic. Points below the reference line of $y = x$ indicate a speedup case with the summary reduction.

We observe in Fig. 7 that the performance of our approach is indeed improved by summary reduction. Among all 445 points, 92 points are near to the reference line, indicating a comparable performance (−5% ~ 5%) with summary reduction; 341 points (including 73 timeout cases) are noticeably below the reference line, indicating a speedup of over 5%; only 12 points are noticeably above the reference line, indicating a slow down of over 5%.

### 7.3 Impact of Lazy Counterexample Analysis

The third experiment tests the impact of lazy counterexample analysis to our regression verification technique. Various results, including the verification time, refinement time and counterexample length are used to evaluate the lazy counterexample analysis.

These results are presented at Fig 8, where Fig 8(a), Fig 8(b) and Fig 8(c) show results on verification time, refinement time and counterexample length, respectively. Note that the counterexample length is the summation length of counterexamples in all iterations of CEGAR. All results are presented in scatter diagrams, where each point represents a single revision of a device driver. The *X* and *Y* axes indicate the performance with and without lazy counterexample analysis , respectively. Both *X* and *Y* axes are logarithmic. Again, a point above the reference line indicates a case of performance improvement.

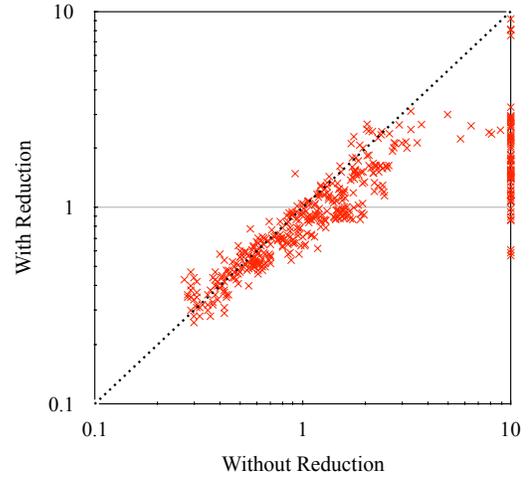

**Figure 7: Compare the total time of summary reuse with-/without summary reduction**

On the whole, among all 445 regression verification tasks, lazy counterexample analysis shows a noticeably speedup (> 5%) in 312 cases for total verification time and in 376 cases for refinement time. It is also observed that this technique yields shorter counterexample paths for the majority (430 out of 445) of cases. These results conform to our algorithmic analysis in Section 6. With the lazy analysis technique, the counterexample needs not be fully expanded. And a shorter counterexample can usually reduce the refinement efforts.

## 8 RELATED WORK

Regression verification was investigated mainly in two directions, the verification of differences, and the reuse of previously computed results.

*Verification of Differences.* In this line of research, one attempts to establish the correctness of the new program by proving its (conditional) equivalence to an old and verified program.

Many techniques have been proposed in this line of research. Golden et al. [19] proposed an technique for proving conditional equivalence of two programs by abstraction and decomposition of procedures. Backes et al. [2] proposed to distinguish the program behaviors that are impacted by the changes. Only the impacted program behaviors needed to be considered during the regression verification. Beyer et al. [7] proposed the conditional model checking, which outputs a condition such that the program satisfies the specification under this condition. Böhme et al. [11] proposed a partition-based regression verification technique. Instead of proving the absence of regression errors for the complete input space, this approach continuously verifies the input space in a gradual

Table 1: Overall experimental results

| Run set | | | | Standalone | | | Summary Reuse | | | | |
|---|---|---|---|---|---|---|---|---|---|---|---|
| Device Driver | Rst | Loc | $T_{1st}$ | $T_{Ref}$ | $T_{Total}$ | #Ref | $T_{Ref}$ | $T_{Total}$ | #Ref | MaxFSize | Speedup |
| mtd-docprobe | 5/0 | 3145 | 3.2 | 1.7 | 15.3 | 55 | 0.3 | 1.9 | 5 | 194.3 | 87.5% |
| block-paride-frpw | 5/0 | 8044 | 21.7 | 33.6 | 102.5 | 95 | 7.0 | 40.1 | 40 | 774.0 | 60.8% |
| media-mt2266 | 5/0 | 4853 | 5.1 | 1.6 | 25.2 | 40 | 0.9 | 10.8 | 35 | 126.3 | 57.0% |
| usb-dwc3-pci | 5/0 | 2784 | 1.2 | 0.7 | 5.7 | 20 | 0.2 | 2.7 | 10 | 55.2 | 52.0% |
| media-max2165 | 5/0 | 5423 | 4.5 | 3.0 | 22.4 | 50 | 1.1 | 10.9 | 40 | 201.1 | 51.2% |
| gpio-74x164 | 5/0 | 3054 | 1.4 | 0.9 | 6.5 | 20 | 0.3 | 3.3 | 15 | 88.5 | 49.8% |
| usb-otg-nop-xceiv | 5/0 | 2075 | 1.4 | 1.0 | 7.4 | 25 | 0.3 | 3.7 | 15 | 71.3 | 49.7% |
| input-ad714x-spi | 5/0 | 3875 | 1.0 | 0.6 | 4.8 | 20 | 0.2 | 2.5 | 10 | 80.7 | 48.3% |
| media-mc44s803 | 5/0 | 4624 | 2.4 | 1.0 | 12.0 | 50 | 0.6 | 6.3 | 25 | 132.9 | 47.9% |
| rtc-ds3234 | 5/0 | 3916 | 1.9 | 1.1 | 9.6 | 20 | 0.5 | 5.1 | 20 | 83.3 | 46.6% |
| hwmon-max1111 | 5/0 | 2986 | 1.4 | 1.0 | 6.8 | 20 | 0.4 | 3.7 | 15 | 85.3 | 45.2% |
| mfd-wl1273-core | 5/0 | 3665 | 1.3 | 0.6 | 6.5 | 10 | 0.3 | 3.6 | 10 | 133.5 | 44.3% |
| addac-adt7316-spi | 4/1 | 2890 | 2.0 | 1.1 | 8.4 | 17 | 0.5 | 4.7 | 23 | 69.7 | 44.1% |
| pci-ioapic | 5/0 | 3678 | 1.4 | 0.8 | 6.8 | 30 | 0.3 | 3.9 | 15 | 40.7 | 43.1% |
| input-adxl34x-spi | 5/0 | 3666 | 1.6 | 0.9 | 7.9 | 35 | 0.4 | 4.6 | 25 | 72.6 | 42.2% |
| comedi-adl_pci7432 | 5/0 | 2720 | 2.0 | 0.9 | 10.3 | 30 | 0.5 | 6.0 | 23 | 118.5 | 41.8% |
| meter-ade7854-i2c | 5/0 | 3805 | 0.7 | 0.6 | 3.6 | 10 | 0.2 | 2.1 | 10 | 170.4 | 41.3% |
| media-gspca_pac207 | 5/0 | 5668 | 3.2 | 1.2 | 16.1 | 55 | 0.8 | 9.5 | 45 | 153.5 | 40.8% |
| rtc-max6902 | 5/0 | 3855 | 1.7 | 1.1 | 8.7 | 20 | 0.6 | 5.3 | 20 | 78.1 | 38.8% |
| dds-ad9910 | 5/0 | 3609 | 1.6 | 1.1 | 8.3 | 30 | 0.5 | 5.3 | 14 | 140.8 | 36.4% |
| gpio-mc33880 | 5/0 | 2953 | 1.3 | 0.8 | 6.6 | 20 | 0.4 | 4.2 | 15 | 78.0 | 36.1% |
| rtc-m41t94 | 5/0 | 4242 | 2.1 | 1.1 | 10.2 | 20 | 0.7 | 6.6 | 20 | 111.8 | 35.9% |
| parport_cs | 5/0 | 4510 | 2.0 | 0.9 | 9.9 | 35 | 0.6 | 6.4 | 25 | 94.4 | 35.6% |
| touchscreen-ad7879 | 5/0 | 3778 | 2.0 | 1.1 | 9.7 | 40 | 0.6 | 6.2 | 25 | 77.9 | 35.5% |
| rtc-stk17ta8 | 5/0 | 4220 | 3.3 | 1.5 | 16.8 | 74 | 1.0 | 10.9 | 70 | 238.3 | 35.2% |
| mtd_pagetest | 5/0 | 4723 | 1.3 | 0.3 | 6.9 | 15 | 0.4 | 4.6 | 15 | 36.1 | 34.3% |
| misc-bmp085 | 5/0 | 3675 | 1.4 | 1.0 | 7.6 | 15 | 0.6 | 5.2 | 18 | 154.7 | 31.1% |
| hid-kensington | 5/0 | 2305 | 0.8 | 0.6 | 4.2 | 20 | 0.2 | 2.9 | 15 | 33.1 | 30.9% |
| mtd_subpagetest | 5/0 | 4483 | 1.3 | 0.3 | 6.4 | 15 | 0.4 | 4.5 | 15 | 268.2 | 30.5% |
| hid-ezkey | 5/0 | 2429 | 1.1 | 0.7 | 5.9 | 29 | 0.3 | 4.2 | 20 | 45.6 | 28.3% |
| staging-zram | 5/0 | 8115 | 0.7 | 0.5 | 3.2 | 5 | 0.2 | 2.3 | 5 | 418.1 | 27.3% |
| comedi-adl_pci7230 | 5/0 | 2661 | 2.1 | 0.9 | 10.4 | 25 | 0.7 | 7.6 | 24 | 111.4 | 27.1% |
| dds-ad9951 | 1/4 | 3088 | 1.6 | 0.9 | 6.0 | 18 | 0.5 | 4.4 | 13 | 82.0 | 25.7% |
| misc-ti_dac7512 | 5/0 | 2623 | 0.4 | 0.4 | 2.2 | 10 | 0.2 | 1.7 | 10 | 47.3 | 24.3% |
| usb-ums-usbat | 5/0 | 9025 | 0.8 | 0.2 | 3.6 | 10 | 0.2 | 2.8 | 10 | 469.1 | 23.8% |
| ⋮ | | | | ⋮ | | | ⋮ | | | | |
| Device Driver | Rst | Loc | $T_{1st}$ | $T_{Ref}$ | $T_{Total}$ | #Ref | $T_{Ref}$ | $T_{Total}$ | #Ref | MaxFSize | Speedup |
| net-phy-realtek | 3/2 | 3528 | 0.4 | 0.4 | 2.3 | 8 | 0.3 | 2.6 | 8 | 23.0 | -13.8% |
| x86-mxm-wmi | 4/1 | 2903 | 0.4 | 0.2 | 2.0 | 10 | 0.2 | 2.3 | 10 | 27.2 | -17.2% |
| net-phy-vitesse | 5/0 | 3704 | 0.5 | 0.4 | 2.5 | 10 | 0.2 | 2.9 | 10 | 32.0 | -18.8% |
| w1_smem | 5/0 | 1072 | 0.3 | 0.3 | 2.0 | 15 | 0.2 | 2.5 | 15 | 18.4 | -27.8% |
| kfifo_buf | 5/0 | 1928 | 0.2 | 0.1 | 0.8 | 5 | 0.1 | 1.6 | 5 | 57.6 | -105.2% |
| **Total** | 418/27 | 342427 | 137.0 | 97.4 | 670.4 | 2175 | 42.7 | 460.2 | 1813 | 774.0 | 31.4% |

manner. Felsing et al. [18] reduced the equivalence proving of two related imperative integer programs to Horn constraints over uninterpreted predicates, and then solved the constraints using an SMT solver. Moreover, Chaki et al. [14] studied the regression verification for multi-threaded programs,

*Reuse of Intermediate Results.* In this line of research, one studies the reuse of previously-generated results to the current verification. A variety of information have been proposed for reuse.

Some researchers[23, 27, 36] proposed to keep the reached state space and reuse them in the further verification runs. The rationale of these techniques is that state spaces of consecutive versions tend to be similar. However, recording and reusing reached state space may be costly, and these techniques may not be applicable to large-scale programs.

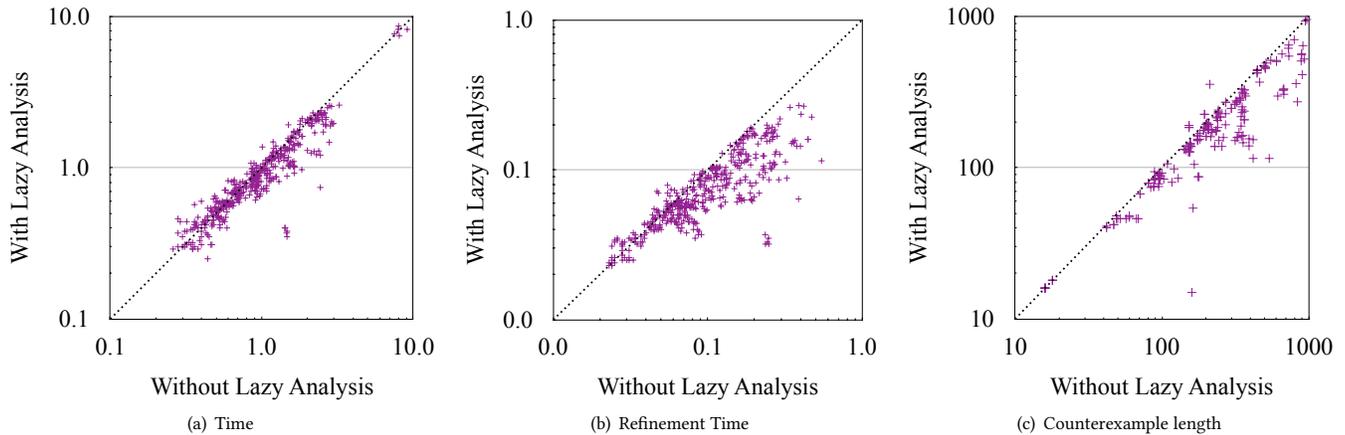

Figure 8: Compare the performance of summary reuse with and without *lazy counterexample analysis*

Visser et al.[35] noticed the importance of constraint solving for symbolic execution. They proposed to cache and reuse the results of constraint solving. This approach was further improved in [1, 25] from different aspects. This group of techniques are orthogonal to our approach. These techniques can be applied to enhance our approach.

Beyer et al.[10] proposed to use abstract precisions as the intermediate results. An abstract precision defines the level of abstraction, which conveys important information on the current verification. They proposed to record the final abstract precision, and to reuse it as the initial abstract precision of the current verification. With this technique, the number of refinements can often be reduced. Note that the precision reuse and our summary reuse are orthogonal to each other. It is possible to combine these two reuse techniques together.

The most relevant work to ours is [17, 32], where an regression verification technique by means of interpolation-based procedure summaries was proposed. However, the definition of procedure summaries is different in that paper. A procedure summary in [32] is an over-approximation of all behaviors of the procedure. In contrast, a procedure summary in our paper is a pair of input and output states, representing a set of visited paths of the procedure. Our definition of summaries is consistent to the classical interprocedural analysis techniques [30, 33], which has been realized in most of the modern software verifiers (for example, CPAchecker). Our summaries can be automatically generated by these interprocedural analysis techniques, without any additional computation effort. In contrast, the interpolation-based summaries in [32] must be computed by an additional process, which in some cases maybe quite time-consuming. On the other hand, the technique in [32] is based on bounded model checking, while ours is developed in the context of CEGAR.

Pastore et al. [29] proposed a method to validate that an already tested code has not been broken by an upgrade. It maintains a test suite that can be used to revalidate the software as it evolves. Different from our approach, this technique is respect to regression testing. The verification technique is used there, as an aid, to validate dynamic properties (or invariants). In contrast, we aim to providing a new regression verification technique via reusing procedure summaries.

## 9 CONCLUSION

We proposed in this paper a fully automatic regression verification technique in the context of CEGAR. Procedure summaries are reused across different abstract precisions and different program revisions. We elaborated techniques for representing, processing and reusing procedure summaries. A summary reduction technique was also proposed to improve the hit rate of summary cache. A lazy counterexample analysis algorithm was further proposed to reduce the unnecessary path expansion efforts. Experimental results show promising performance of our technique.

In the future, we are planning to investigate other kinds of intermediate results that can be reused in regression verification.